\documentclass[fleqn,usenatbib]{mnras}
\usepackage{amsmath}
\usepackage{graphicx}
\usepackage{amssymb,txfonts}

\newcommand{\msun}{{\rm M}_{\sun}}

\newbox\grsign \setbox\grsign=\hbox{$>$} \newdimen\grdimen \grdimen=\ht\grsign
\newbox\simpropbox
\setbox\simpropbox=\hbox{\raise.5ex\hbox{$\propto$}\llap
   {\lower.5ex\hbox{$\sim$}}}\ht2=\grdimen\dp2=0pt
\def\simprop{\mathrel{\copy\simpropbox}}

\title[Adiabatic losses and soft radio spectra]{The effect of adiabatic losses on spectra of stationary jets and the origin of soft radio spectra of accreting black-hole sources}

\author[A.~A. Zdziarski et al.]
{Andrzej A. Zdziarski,$^1$\thanks{E-mail: aaz@camk.edu.pl (AAZ), stawarz@oa.uj.edu.pl ({\L}S), sikora@camk.edu.pl (MS).} {\L}ukasz Stawarz$^2$\footnotemark[1] and Marek Sikora$^1$\footnotemark[1]\\
$^1$Nicolaus Copernicus Astronomical Center, Polish Academy of Sciences, Bartycka 18, PL-00-716 Warszawa, Poland\\
$^2$Astronomical Observatory, Jagiellonian University, Orla 171, PL-30-244 Krak{\'o}w, Poland
}

\date{Accepted 2019 February 12. Received 2019 February 12; in original form 2018 December 29}

\pagerange{\pageref{firstpage}--\pageref{lastpage}}
\pubyear{2019}

\begin{document}

\maketitle

\label{firstpage}

\begin{abstract}
It has been suggested that adiabatic energy losses are not effective in stationary jets, where the jet expansion is not associated with net work. Here, we study jet solutions without them, assuming that adiabatic losses are balanced by electron reacceleration. The absence of effective adiabatic losses makes electron advection along the jet an important process, and we solve the electron kinetic equation including that process. We find analytical solutions for the case of conical jets with advection and synchrotron losses. We show that accounting for adiabatic losses in the case of sources showing soft partially self-absorbed spectra with the spectral index of $\alpha<0$ in the radio-to-IR regime requires deposition of large amounts of energy at large distances in the jet. On the other hand, such spectra can be accounted for by advection of electrons in the jet. We compare our results to the quiescent spectrum of the blazar Mrk 421. We find its soft radio-IR spectrum can be fitted either by a model without adiabatic losses and advection of electrons or by one with adiabatic losses, but the latter requires injection of a very large power at large distances.
\end{abstract}
\begin{keywords}
acceleration of particles -- BL Lacertae objects: individual (Mrk~421) -- radiation mechanisms: non-thermal -- galaxies: active -- galaxies: jets -- X-rays: binaries.
\end{keywords}

\section{Introduction}
\label{intro}

The nature of adiabatic losses in stationary jets has been a subject of a dispute. If the jet radius is increasing with the distance from the origin, relativistic electrons moving along it expand with the jet and are subject of adiabatic losses. On the other hand, a stationary jet has a shape constant in time, and performs no work on the external medium. Then, the energy lost adiabatically has to be converted into another form, e.g., can be used for a secondary bulk acceleration of the jet \citep{laing04}. 

Adiabatic losses are usually the dominant ones in black-hole (BH) jets at large distances from the jet origin, as the magnetic and photon fields are typically weak there. In the simplest approach, e.g., \citet{bk79} (hereafter BK79), \citet{konigl81}, \citet*{zls12}, the form of the electron distribution is assumed along the jet. However, if the losses are present and the electron distribution integrated over the jet cross section remains constant along the jet (as in the model of BK79), there has to be a continuous acceleration of electrons within the jet in order to compensate the losses. For example, \citet{jester01,jester05} found that continuous electron acceleration is required in the quasar 3C 273. 

The model of BK79 was developed in order to account for flat radio spectra, $F(\nu) \simprop \nu^\alpha$ with $\alpha\sim 0$ [where $F(\nu)$ is the energy flux per unit photon frequency], which are common in both the cores of flat-spectrum radio quasars (e.g., \citealt{yuan18}) and the hard state of BH binaries (e.g., \citealt{fender00}). Such spectra appear due to a superposition of a synchrotron self-absorbed emission up to some distance along the jet, and an optically thick emission above this distance. The characteristic distance from the jet origin at which most of the flux at a frequency $\nu$ is emitted scales as $z\propto \nu^{-1}$. As the flat radio spectra can extend down to $\sim$100 MHz (e.g., \citealt{zywucka14}), this implies the emission of low-frequency radiation at rather large distances. 

In steady state, the spectral index of partially synchrotron self-absorbed emission depends on the spatial distributions of both the relativistic electrons and magnetic fields. If these distributions are given by power laws of the distance, see equation (\ref{pl}) in the Appendix \ref{approx}, the spectral index is a function of the two spatial indices. We give this dependence in equation (\ref{alpha}). For a conserved electron number at a given energy and conserved magnetic energy flux in a conical jet with a constant speed, the two corresponding indices are $a=2$ and $b=1$, respectively, in which case $\alpha=0$. If either the electrons suffer radiative and adiabatic losses or the magnetic energy is dissipated, $a>2$, $b>1$, respectively. Equation (\ref{alpha}) shows that in either case the spectrum becomes harder, $\alpha>0$. Thus, spectra harder than $\alpha=0$ appear naturally in the presence of dissipation of the jet internal energy. On the other hand, spectra softer than $\alpha=0$ require the energy fluxes of either electrons or magnetic field to increase along the jet. The latter would require an external source of magnetic flux, which appears unlikely. Thus, spectra with $\alpha<0$ require deposition of substantial amounts of energy in relativistic electrons at large distances, which energy may then be lost mostly adiabatically. Soft spectra with $\alpha<0$ are found to be emitted by radio cores of about a half of the radio-loud AGN sample analysed by \citet{yuan18}.

An example of a blazar with a soft radio spectrum is Mrk 421, which has $\alpha\simeq -0.2$ in its quiescent state \citep{abdo11b}. If this spectrum is due to continuous electron acceleration and adiabatic losses, the power lost adiabatically is estimated in this work to be $\sim\! 10^{45}$ erg s$^{-1}$, much larger than the estimated jet bolometric luminosity of $\sim\! 10^{43}$ erg s$^{-1}$. Even if the power lost adiabatically is converted into bulk acceleration of the jet \citep{laing04}, it still would need to be supplied to the jet, imposing severe requirements on the energetics of that AGN. 

In order to mitigate the problem of the large adiabatic losses, a number of authors, e.g., \citet{kaiser06}, \citet{pe'er09} and \citet{potter12, potter13a, potter13b} considered the presence of adiabatic losses as optional, and studied jet models both with and without adiabatic losses. Some others developed jet models with acceleration and radiative losses but without including adiabatic losses at all, e.g., \citet*{hervet15}. The issue of the nature of adiabatic losses is discussed, in particular, in \citet{potter12,potter15}, who argued that both the bulk velocity and the electron internal energy may remain constant in a conical jet of constant speed. \citet{potter15} considered the possibility that adiabatic losses are converted into radial bulk motion, which is then immediately converted back into electron acceleration.

Given the uncertain theoretical status of adiabatic losses in a steady-state jet, we study here solutions without them. This case corresponds to the adiabatic energy loss rate being balanced by the electron reacceleration rate,
\begin{equation}
-\dot\gamma_{\rm ad}=\dot\gamma_{\rm reacc}.
\label{balance}
\end{equation}
Admittedly, the validity of this equality is uncertain. However, it is the simplest case with the minimum of assumptions. It can be due to, e.g., the adiabatic-loss energy converted into turbulence, which then re-accelerates the electrons in the jet. In this work, we illustrate the effect of the absence of adiabatic losses on the electron and photon distributions. 

However, neglecting the adiabatic losses makes electron advection along the jet an important process, and thus we need to solve the electron kinetic equation including that process. We follow here the method of \citet{z14a} (hereafter Z14), who considered steady solutions of jets with continuous acceleration including adiabatic losses. We obtain analytical solutions for the case of conical jets with advection and synchrotron losses. These solutions supplement the classical ones of \citet{kardashev62}. We find that the electron advection along the jet in the absence of effective adiabatic losses leads to accumulation of advected electrons increasing with the distance, which can then fit the synchrotron spectrum observed from Mrk 421.

We note that the approach we use in our study is different from those employed by \citet{kaiser06} or \citet{pe'er09}, who considered a single acceleration episode and subsequent propagation of the electrons. Here, we find a continuous acceleration is necessary to account for the observed spectra with $\alpha<0$. Our approach is also different from that of \citet{potter12, potter13a, potter13b, potter15}, who considered numerical jet models. Instead, we here model the jet using a kinetic equation and obtain analytical solutions. Electron advection in jets has also been recently considered by \citet*{khangulyan18}, but in a different context and using a different method from ours.

In Section \ref{no_adiabatic}, we set up the problem and provide solutions to the case without adiabatic losses. In Section \ref{adiabatic}, we generalize the previous solution of Z14 for the case with adiabatic losses. In Section \ref{comparison}, we compare our models to the quiescent spectrum from radio to X-rays of Mrk 421. We discuss and summarize our results in Section \ref{discussion}. In Appendix \ref{approx}, we provide formulae for the synchrotron flux with an arbitrary dependence of the radius and velocity on the distance and present the results for the case of the steady-state electron distribution and magnetic field being each a power law in the distance.

\section{Jet solutions without adiabatic losses}
\label{no_adiabatic}

We follow here the notation of Z14 (with small modifications). The electron kinetic equation for an axially symmetric jet with arbitrary profiles of the radius, $r_{\rm j}(z)$, velocity and energy loss rate but neglecting diffusion terms is given by equation 24 of Z14. Following that work, we introduce a dimensionless distance, $\xi\equiv z/z_0$, and a dimensionless energy loss rate,
\begin{equation}
\tilde{\gamma}\equiv \dot \gamma\frac{{\rm d} t}{{\rm d}\xi}=
\frac{\dot \gamma z_0}{c \Gamma\beta_{\rm j}(\xi)},
\label{gdot_xi}
\end{equation}
where $z$ is the distance along the jet measured from the BH centre, $z_0$ is the distance at which the electron acceleration begins, $\gamma$ and $\Gamma$ are the electron and bulk Lorentz factors, respectively, and $\beta_{\rm j} c$ is the bulk jet velocity. Also, we define dimensionless distributions integrated over the jet cross section, 
\begin{align}
&\tilde{Q}(\gamma,\xi)\equiv {\upi r_{\rm j}(\xi)^2 z_0^2\over c} Q(\gamma,\xi),
\label{Q}\\
&\tilde{N}(\gamma,\xi)\equiv \upi\Gamma\beta_{\rm j}(\xi)z_0 r_{\rm j}(\xi)^2 N(\gamma,\xi),\label{N}
\end{align}
where $Q$ is the rate of injection of relativistic electrons per unit volume and $N$ is the steady-state electron distribution. Here, we have neglected possible radial inhomogeneity of the jet, in particular clumping and/or a spine/sheath structure. The dimensionless kinetic equation (neglecting dispersion terms) becomes then,
\begin{equation}
{\partial\over \partial \xi}\tilde{N}(\gamma,\xi)+{\partial\over \partial \gamma}\left[\tilde{\gamma} \tilde{N}(\gamma,\xi)\right] =\tilde{Q}(\gamma,\xi)
\label{kinetic}
\end{equation}
(as given by equation 27 of Z14). This equation has two limiting solutions. If the spatial advection term is negligible, we have the usual cooling-dominated solution,
\begin{equation}
\tilde{N}(\gamma,\xi)=\frac{1}{\tilde{\gamma}}\int_\gamma^\infty {\rm d}\gamma' \tilde{Q}(\gamma',\xi).
\label{cooling}
\end{equation}
If the cooling term is negligible, the solution with the boundary condition of $\tilde{N}(\gamma,1)=0$ is
\begin{equation}
\tilde{N}(\gamma,\xi)=\int_1^\xi {\rm d}\xi' \tilde{Q}(\gamma,\xi'),
\label{advection}
\end{equation}
which corresponds to the accumulation at a given $\xi$ of the electrons produced below it.

The general solution of equation (\ref{kinetic}) is given by equation 29 of Z14, derived based on \citet{stawarz08}. In the case of only the optically-thin synchrotron and Thomson losses and without adiabatic losses, we find the solution to be,
\begin{equation}
\tilde{N}(\gamma,\xi)=\frac{1}{\gamma^{2}}\int_{\xi_{\rm m}(\gamma,\xi)}^\xi \!\!\!\!{\rm d}\xi' \tilde{Q}[\gamma'(\gamma,\xi;\xi'),\xi'] \gamma'(\gamma,\xi;\xi')^{2}.
\label{eldist}
\end{equation}
Here $\gamma'(\gamma,\xi;\xi')>\gamma$ is the Lorentz factor at $\xi'<\xi$ given its value of $\gamma$ at $\xi$ (cf.\ \citealt{kaiser06,pe'er09}),
\begin{align} 
&\gamma'(\gamma,\xi;\xi')={\gamma \over 1- \gamma \int_{\xi'}^{\xi} {\rm d}\xi'' c_1(\xi'')U(\xi'') },\label{gamma_xi0}\\
&c_1(\xi)\equiv \frac{4\sigma_{\rm T}z_0}{ 3 m_{\rm e}c^2\Gamma\beta_{\rm j}(\xi)},
\end{align}
where $U$ is the energy density of both magnetic field and radiation, $m_{\rm e}$ is the electron mass. The dimensionless distance, $\xi_{\rm m}$, is given by the maximum of 1 and the solution of
\begin{equation}
\int_{\xi_{\rm m}(\gamma,\xi)}^{\xi} {\rm d}\xi' c_1(\xi')U(\xi') =\gamma^{-1},
\label{xm0}
\end{equation}
and represents the minimum distance along the jet from which an electron with the Lorentz factor $\gamma$ located at $\xi$ can be advected.

We then consider the special case of a conical jet with constant speed and conserved magnetic energy flux, i.e., 
\begin{equation}
r_{\rm j}(\xi)=\xi z_0\tan\Theta,\quad B(\xi)=B_0 \xi^{-1},
\label{r_B}
\end{equation}
where $\Theta$ is the half-opening angle. We also assume a power-law rate in both the electron Lorentz factor and distance, i.e.,
\begin{equation}
\tilde{Q}=\tilde{Q}_0 \xi^{-k} \gamma^{-p},\quad \gamma_{\rm min}\leq \gamma
\leq\gamma_{\rm max}(\xi),\quad 1\leq \xi \leq \xi_{\rm max},
\label{Q1}
\end{equation}
where $\gamma_{\rm min}$ and $\gamma_{\rm max}$ give the range within which the electrons are accelerated, $\xi_{\rm max}=z_{\rm max}/z_0$, $z_{\rm max}$ is the jet distance at which the injection of fresh relativistic particles ceases, and $p$ and $k$ are the electron acceleration and injection rate indices, respectively. 

We assume a magnetic acceleration rate of electrons as $\dot \gamma_{\rm acc} m_{\rm e} c=\eta_{\rm acc} e B$, where $m_{\rm e}$ and $e$ are the electron mass and charge, respectively, balanced by synchrotron losses (as in Z14), where the efficiency factor $\eta_{\rm acc}$ is usually $\la 1/2\upi$. This yields
\begin{equation}
\gamma_{\rm max}(\xi) \simeq \left[9 \eta_{\rm acc}B_{\rm cr} \over 4\alpha_{\rm f} B(\xi)\right]^{1/2}=\gamma_{\rm max0}\xi^{1/2},\quad \gamma_{\rm max0}=
\left(9 \eta_{\rm acc}B_{\rm cr}\over 4\alpha_{\rm f} B_0\right)^{1/2},
\label{gemax}
\end{equation}
where $\alpha_{\rm f}$ is the fine-structure constant and $B_{\rm cr}$ is the critical magnetic field. It is important that then the corresponding maximum photon energy, $\propto \gamma_{\rm max}^2 B$, is independent of $B$ \citep*{gfr83}. We note that the jet continues to propagate beyond $z_{\rm max}$, which regime we consider below under the same assumptions as at $<z_{\rm max}$.

We then assume optically-thin synchrotron losses and $\gamma\gg 1$, for which the loss rate is
\begin{equation}
\tilde{\gamma}_{\rm s}(\gamma,\xi)=- q \frac{\gamma^2}{\xi^2}
,\quad q\equiv \frac{\sigma_{\rm T}z_0 B_0^2}{ 6\upi m_{\rm e}c^2\Gamma\beta_{\rm j}},
\label{gamma_dot_s}
\end{equation}
where $q$ characterizes the ratio of the rate of the synchrotron losses to that of the advection. Then we find,
\begin{align}
&\gamma'(\gamma,\xi;\xi')=\frac{\gamma}{1-q\gamma(1/\xi'-1/\xi)},
\label{gamma_prime}\\
&\xi_{\rm m}(\xi)= \begin{cases}1, &\gamma\leq \gamma_{\rm adv}(\xi);\cr
\xi\left(1+\frac{\xi}{q\gamma}\right)^{-1}, &\gamma> \gamma_{\rm adv}(\xi),\cr
\end{cases}\label{xim}\\
&\gamma_{\rm adv}(\xi)\equiv \frac{\xi}{q(\xi-1)},
\label{gadv}
\end{align}
where $\gamma_{\rm adv}(\xi)$ is the Lorentz factor at which there is a break in the solutions due to the presence of the boundary of the acceleration region at $\xi=1$ (note that its value is different from that for the solution including adiabatic losses, cf.\ equation 37 in Z14). Then, we find the solution for $\gamma_{\rm max}\rightarrow \infty$ as
\begin{equation}
\tilde{N}(\gamma,\xi)=
\tilde{Q}_0\gamma^{-2} \int_{\xi_{\rm m}}^\xi {\rm d}\xi' {\xi'}^{-k} \gamma'(\gamma,\xi;\xi')^{2-p},\label{N_analytical}
\end{equation}
which for $k\neq 1$ yields,
\begin{align}
&\tilde{N}(\gamma,\xi)=
\tilde{Q}_0\gamma^{-p}\times\label{Nkneq1}\\
&\left.\! \frac{ (\xi\xi')^{2-p}{\xi'}^{-k} \left[\xi\xi'+q\gamma(\xi'-\xi)\right]^p\!
{_2F_1}\!\left(1,-2\!+\!k\!+\!p,k;\!\frac{q\gamma\xi}{q\gamma\xi'+\xi\xi'}\right)}{(k-1)(q \gamma+\xi)\left[q\gamma\xi -(q\gamma+\xi)\xi'\right]}
\right|^{\xi'=\xi}_{\xi'=\xi_{\rm m}},\nonumber
\end{align}
where ${_2}F_1$ is the Gauss hypergeometric function. For $k=1$ and non-integer $p$, we find,
\begin{align}
&\tilde{N}(\gamma,\xi)=
\tilde{Q}_0\gamma^{-1-p}\times\label{Nkeq1}\\
&\left.\! \frac{ (\xi\xi')^{1-p}{\xi'} \left[\xi\xi'+q\gamma(\xi'-\xi)\right]^{p-1}\!
{_2F_1}\!\left(1,1,3-p; \frac{\xi'}{q\gamma} +\frac{\xi'}{\xi}\right)}{q (p-2)}\right|^{\xi'=\xi}_{\xi'=\xi_{\rm m}}.\nonumber
\end{align}
For $k=1$, $p=2$, we have a simple solution,
\begin{equation}
\tilde{N}(\gamma,\xi)=\frac{\tilde{Q}_0}{\gamma^2} \begin{cases}\ln\xi, &\gamma\leq \gamma_{\rm adv}(\xi);\cr
\ln\left(1+\frac{\xi}{q\gamma}\right), &\gamma> \gamma_{\rm adv}(\xi);\cr
\frac{\xi} {q\gamma}, &\gamma\gg \xi/q,\cr
\end{cases}
\label{N2}
\end{equation}
where the last equality is approximate. The independence of the solution for $\gamma<\gamma_{\rm adv}$ of $B_0$ is due to the middle term in equation (\ref{kinetic}) being null in that case. Note that now the normalization of the $\gamma$-dependent part of the dimensionless electron distribution increases with the distance, $\propto\ln \xi$ at low $\gamma$ and $\propto \xi$ at large $\gamma$. For $k=1$, $p=3$, we have
\begin{equation}
\tilde{N}(\gamma,\xi)=\frac{\tilde{Q}_0}{\gamma^3} \begin{cases}\left[ \frac{(1-\xi)q\gamma}{\xi}+\left(1+\frac{q\gamma}{\xi}\right)\ln\xi \right], &\gamma\leq \gamma_{\rm adv}(\xi);\cr
\left[\left(1+\frac{q\gamma}{\xi} \right) \ln\left(1+\frac{\xi}{q\gamma}\right) -1\right], &\gamma> \gamma_{\rm adv}(\xi);\cr
\frac{\xi} {2 a\gamma}, &\gamma\gg \xi/q.\cr
\end{cases}
\label{N3}
\end{equation}

\begin{figure}
\centerline{\includegraphics[height=0.95\columnwidth]{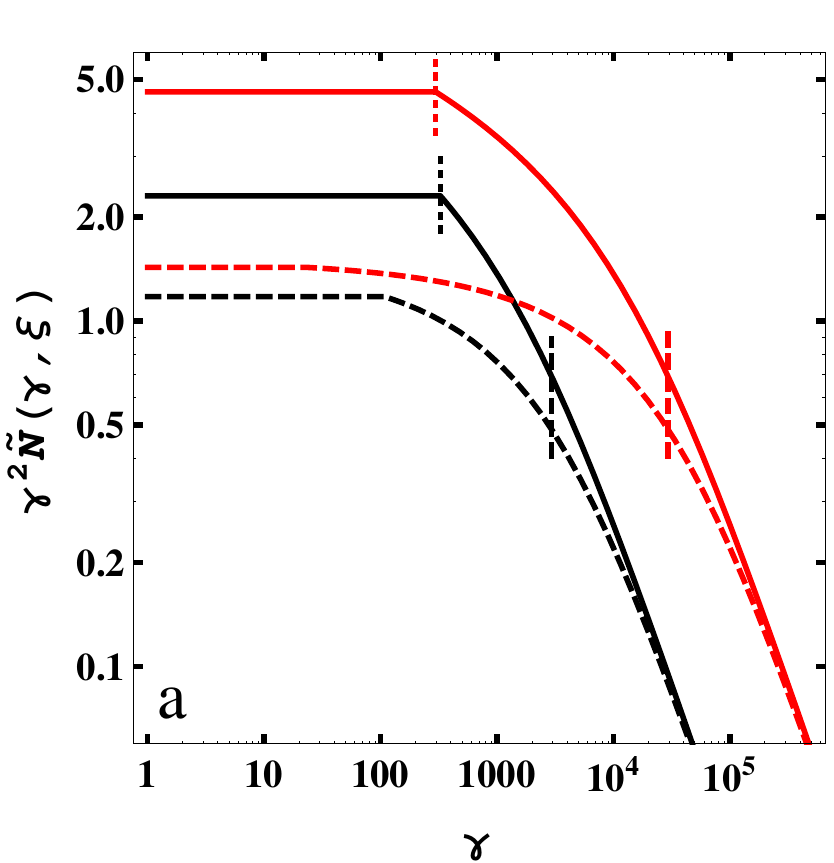}} \centerline{\includegraphics[height=0.95\columnwidth]{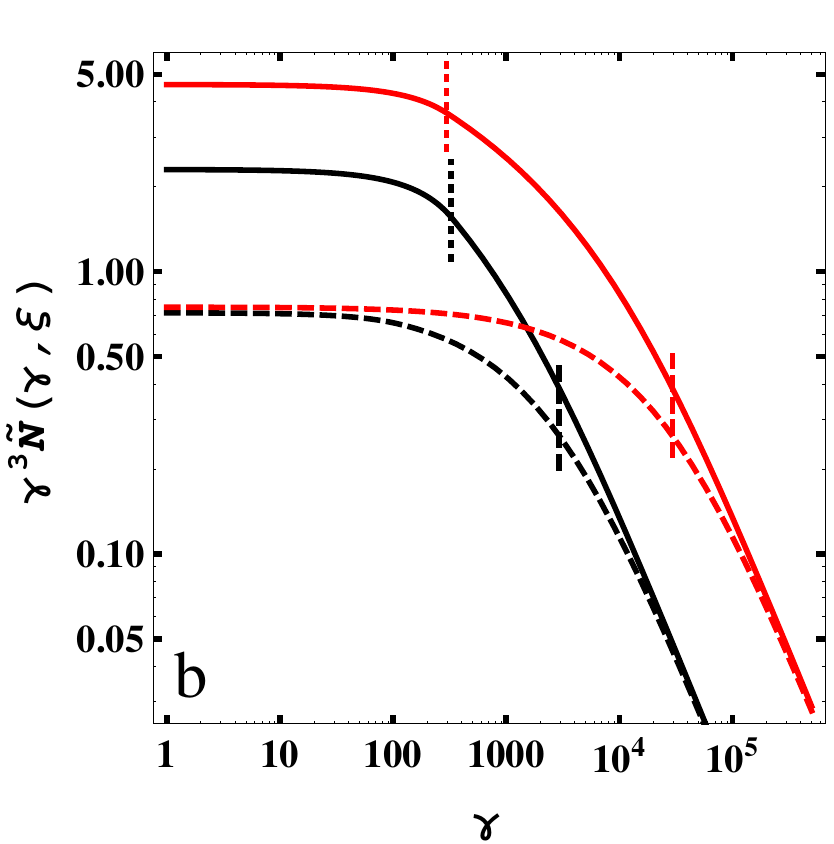}} 
\caption{Comparison of the exact solutions to the advection-losses kinetic equation with only synchrotron losses (solid curves) and with both adiabatic and synchrotron losses (dashed curves; see Z14), for $q=0.0034$ and (a) $p=2$ and (b) $p=3$. The black (lower) and red (upper) curves correspond to $\xi=10$ and 100, respectively. The values of $\gamma_{\rm adv}$ and $\gamma=\xi/q$ are marked with the vertical dotted and dashed lines, respectively. The normalization corresponds to $\tilde{Q}_0=1$.
} \label{p2_3}
\end{figure}

Fig.\ \ref{p2_3} shows some examples of the distributions with advection and radiative losses only, and compares them to the solutions including adiabatic losses. Hereafter, we use the formulae for $\gamma\gg 1$, which is assumed in our solutions, down to $\gamma=1$. Since the emission of electrons with $\gamma\sim 1$ gives a minor contribution to the jet spectra in most cases, the error made by this assumption is small. Still, we should bear in mind the use of this approximation.  We see that the main steepening of the former distribution takes place around $\gamma_{\rm adv}$, and the usual synchrotron break, at $\xi/q$, is not associated with a significant curvature. The distributions at $\gamma\la \gamma_{\rm adv}$ for the case without adiabatic losses are also much above those with those losses. While Z14 found that the solutions with advection and all losses are relatively similar to those with local cooling only, this is not the case in the absence of adiabatic losses. Still, our new solutions show distinct steepenings, asymptotically as $\Delta p=1$. This is because advection along the jet acts on a similar time scale as adiabatic losses, which are also similar to a sink of electrons from a localized region (in which case the first term in equation \ref{kinetic} would be replaced by $\tilde{N}/\xi_{\rm escape}$, where $\xi_{\rm escape}$ is an electron escape scale). However, our solutions are markedly different from those for synchrotron cooling only, which would be given by a steepening of the electron distribution with respect to the injected distribution in the entire energy range, cf.\ equations (\ref{cooling}) and (\ref{gamma_dot_s}). 

In our analysis, we have neglected the effect of the high-energy cutoff of the acceleration rate. However, for the usual range of $p\ga 1.5$, the contribution in number of electrons accelerated at high energies to the steady-state distributions at lower energies is minor (while it may become dominant at $p\leq 1$). Then at $p<2$, while most of the energy is injected at the high-energy end of the acceleration rate, that energy is radiated locally at high energies and contributes negligibly to the radiation at lower energies, due to the single-electron synchrotron spectrum being relatively narrow. Also, the contribution of cooled electrons to the distribution at lower energies is minor. However, we need to take into account the high-energy cutoff in calculating the electron distribution at $\xi>\xi_{\rm max}$, the power injected into a jet, and the radiated photon spectrum.

We now consider the part of the jet at $\xi>\xi_{\rm max}$ under the same assumption as for the accelerating part of the jet, i.e., adiabatic losses balanced by reacceleration. The electron distribution there is given by equation (\ref{eldist}) but with $\tilde{Q}=0$. As the boundary condition we take the electron distribution at $\xi_{\rm max}$,
\begin{equation}
\tilde{Q}(\gamma,\xi_{\rm max})=\tilde{N}\delta_{\rm D}(\xi-\xi_{\rm max}),
\label{Nmax}
\end{equation}
where $\delta_{\rm D}$ is the Dirac delta function. This yields
\begin{equation}
\tilde{N}(\gamma,\xi)= \begin{cases}\displaystyle{\left(\frac{\gamma'}{\gamma} \right)^2 \tilde{N}(\gamma',\xi_{\rm max}),} & \gamma'<\gamma_{\rm max}(\xi_{\rm max});\cr
0, & \gamma'\geq\gamma_{\rm max}(\xi_{\rm max}),\cr
\end{cases}
\label{Nhigh}
\end{equation}
where $\gamma'$ is given by equation (\ref{gamma_prime}) for $\xi'=\xi_{\rm max}$,
\begin{equation} 
\gamma'(\gamma,\xi;\xi_{\rm max})=\frac{\gamma}{1-q\gamma(1/\xi_{\rm max}-1/\xi)},
\label{gamma_prime_high}
\end{equation}
and the condition of $\gamma'<\gamma_{\rm max}(\xi_{\rm max})$ corresponds to the upper limit on $\gamma$ of
\begin{equation}
\gamma_{\rm max}(\xi>\xi_{\rm max})\equiv
\left[\frac{1}{\gamma_{\rm max}(\xi_{\rm max})}+q\left(\frac{1}{\xi_{\rm max}}-\frac{1}{\xi}\right)\right]^{-1}.
\label{gmax_prime}
\end{equation}
Note that under our assumptions, the electrons never completely lose their energy in the synchrotron process (as noted by \citealt{kaiser06}), and $\gamma_{\rm max}(\xi\gg \xi_{\rm max})\simeq \gamma_{\rm max}(\xi_{\rm max})/(1+\gamma_{\rm max}(\xi_{\rm max}) a/\xi_{\rm max})$. The electrons would still lose their energy in other processes, e.g., Compton scattering of the CMB and other radiation fields, and synchrotron emission in the ISM/IGM magnetic fields. The above solution is analogous to that given in the appendix of \citet{z14b} for the case with adiabatic losses. However, the presence of adiabatic losses in that case causes the electrons to quickly lose all of their energy beyond $\xi_{\rm max}$.

We then consider the power supplied to the electrons, the jet power and the energy loss rate. For a constant-speed jet with any shape and relativistic electrons we have the power in the injected electrons as
\begin{align}
&\frac{{\rm d}P_{\rm inj}}{{\rm d}\xi} = \upi z_0\, m_{\rm e} c^2  \Gamma r_{\rm j}(\xi)^2\,\! \int_{\gamma_{\rm min}}^{\gamma_{\rm max}(\xi)} {\rm d}\gamma\, (\gamma-1) Q(\gamma,\xi)\nonumber\\
&\quad\quad\,\,\,=\frac{m_{\rm e}c^3\Gamma}{z_0} \int_{\gamma_{\rm min}}^{\gamma_{\rm max}(\xi)} {\rm d}\gamma\, (\gamma-1) \tilde{Q}(\gamma,\xi),\label{Pinj_xi}\\
&P_{\rm inj}(\xi) =2\int_1^{\xi} {\rm d}\xi'\, \frac{{\rm d}P_{\rm inj}}{{\rm d}\xi'},
\quad P_{\rm inj}\equiv P_{\rm inj}(\xi_{\rm max}),
\label{Pinj}
\end{align}
respectively, where $P_{\rm inj}(\xi)$ is the power injected up to $\xi$, and the factor of 2 accounts for the presence of both the jet and counterjet. For the injection rate of equation (\ref{Q1}) in a conical jet and approximating $\gamma-1\approx \gamma$, we have
\begin{align}
&P_{\rm inj}\approx \frac{2 m_{\rm e}c^3\Gamma}{z_0}\tilde{Q_0}I_{\rm inj},\label{Pinj12} \\
&I_{\rm inj}\equiv 
\begin{cases}\ln\xi_{\rm max} \ln (\gamma_{\rm max0}\,\xi_{\rm max}^{1/4}/ \gamma_{\rm min}),&k=1,\,p=2;\cr
\displaystyle{
\frac{2\gamma_{\rm max0}^{2-p}(\xi_{\rm max}^{1-p/2}-1)}{(2-p)^2}
-\frac{\gamma_{\rm min}^{2-p}\ln \xi_{\rm max}}{2-p}},&k=1,\,p\neq 2;\cr
\frac{1-\xi_{\rm max}^{1-k}}{k-1} \left[\frac{1}{2(k-1)}+ \ln\frac{\gamma_{\rm max0}}{\gamma_{\rm min}}\right]-\frac{\xi_{\rm max}^{1-k}}{k-1}\ln \xi^{1/2},&k\neq 1,\,p=2;\cr
\displaystyle{
\frac{\gamma_{\rm min}^{2-p}(1-z_{\rm max}^{1-k})}{(k-1)(p-2)}
-\frac{\gamma_{\rm max0}^{2-p}(1-z_{\rm max}^{2-k-p/2})}{(k+p/2-2)(p-2)}}
,&k\neq 1,\,p\neq 2.\cr\end{cases}
\label{Iinj}
\end{align}
The power in the bulk motion of the internal energy of electrons and/or pairs at a given $\xi$ is,
\begin{align}
&P_{\rm e}(\xi)=2 \upi  m_{\rm e} c^3  \Gamma^2 \beta_{\rm j} r_{\rm j}(\xi)^2\,\! \int_1^\infty {\rm d}\gamma\, (\gamma-1) N(\gamma,\xi)\nonumber\\
&\quad\quad\,\,\,=\frac{2 m_{\rm e}c^3\Gamma}{z_0} \int_1^\infty {\rm d}\gamma\, (\gamma-1) \tilde{N}(\gamma,\xi)
\label{P_e}\\
&\quad\quad\,\,\,=P_{\rm inj}(\xi) \frac{\int_1^\infty {\rm d}\gamma\, (\gamma-1) \tilde{N}(\gamma,\xi)}{\int_1^\xi {\rm d}\xi' \int_{\gamma_{\rm min}}^{\gamma_{\rm max}(\xi')} {\rm d}\gamma\, (\gamma-1)\tilde{Q}(\gamma,\xi').}\label{P_e_inj}
\end{align}
In the case of no radiative losses, $\tilde{N}$ is given by equation (\ref{advection}), and then equation (\ref{P_e_inj}) implies that $P_{\rm e}(\xi)=P_{\rm inj}(\xi)$, which is a natural result, expressing the fact that the jet power in the electron internal energy is due to the power provided by their acceleration (minus the radiated power). 

The power in the bulk motion of ions (cf., e.g., \citealt{zdz14}), is constant and it can be related to the integral over the distribution of the relativistic electrons at $\xi_{\rm max}$, where that quantity is maximized,
\begin{align}
&P_{\rm i}=2\upi \mu_{\rm pl} m_{\rm p} c^3  \Gamma(\Gamma-1) \beta_{\rm j} r_{\rm j}(\xi_{\rm max})^2\,\! \int_1^\infty {\rm d}\gamma\, N(\gamma,\xi_{\rm max})\nonumber\\
&\quad\quad\,\,\,=\frac{2\mu_{\rm pl} m_{\rm p}c^3(\Gamma-1)}{z_0} \int_1^\infty {\rm d}\gamma\, \tilde{N}(\gamma,\xi_{\rm max})
\label{P_i}
\end{align}
where $\mu_{\rm pl}$ is the mean molecular weight per electron in the non-thermal distribution at $\xi_{\rm max}$ (which equals to the mean ion molecular weight, $\mu_{\rm i}$, if the only electrons are the non-thermal ones and there are no e$^\pm$ pairs), $m_{\rm p}$ is the proton mass, and we assumed the ion internal energy to be negligible. The power in the magnetic field is
\begin{equation}
P_B(\xi)=2 \upi  \Gamma^2 \beta_{\rm j}c r_{\rm j}(\xi)^2 \frac{B(\xi)^2}{4\upi},
\label{P_B}
\end{equation}
which is constant, $\equiv P_B$, for our assumed case of conserved energy flux, equation (\ref{r_B}), and the factor of $4\upi$ in the denominator corresponds to the case of dominant toroidal magnetic field (consistent with that assumption).

The local and integrated energy powers for either a radiative or adiabatic process are
\begin{align}
&\frac{{\rm d}P_{\rm loss}}{{\rm d}\xi} =-\upi z_0\, m_{\rm e} c^2 \Gamma r_{\rm j}(\xi)^2\! \int_1^\infty {\rm d}\gamma\, N(\gamma,\xi) \dot\gamma(\gamma,\xi),\nonumber\\
&\quad\quad\,\,\,=-\frac{m_{\rm e}c^3\Gamma}{z_0} \int_1^\infty {\rm d}\gamma\,\tilde{N}(\gamma,\xi)\tilde{\gamma}(\gamma,\xi)
\label{Prad_xi}\\
&P_{\rm loss}(\xi) =2\int_1^\xi {\rm d}\xi'\, \frac{{\rm d}P_{\rm loss}}{{\rm d}\xi'},\quad P_{\rm loss}\equiv P_{\rm loss}(\infty),
\label{Prad}
\end{align}
respectively. In the present case, $\tilde{\gamma}$ corresponds to the synchrotron process. In the case of fully efficient cooling, equation (\ref{cooling}), the integral in equation (\ref{Prad_xi}) becomes (using integration by parts) $\int (\gamma-1)\tilde{Q}(\gamma, \xi)$, and then ${\rm d}P_{\rm loss}/{\rm d}\xi={\rm d}P_{\rm inj}/{\rm d}\xi$, i.e., the power injected at a given $\xi$ equals to the lost power.

\begin{figure}
\centerline{\includegraphics[height=0.95\columnwidth]{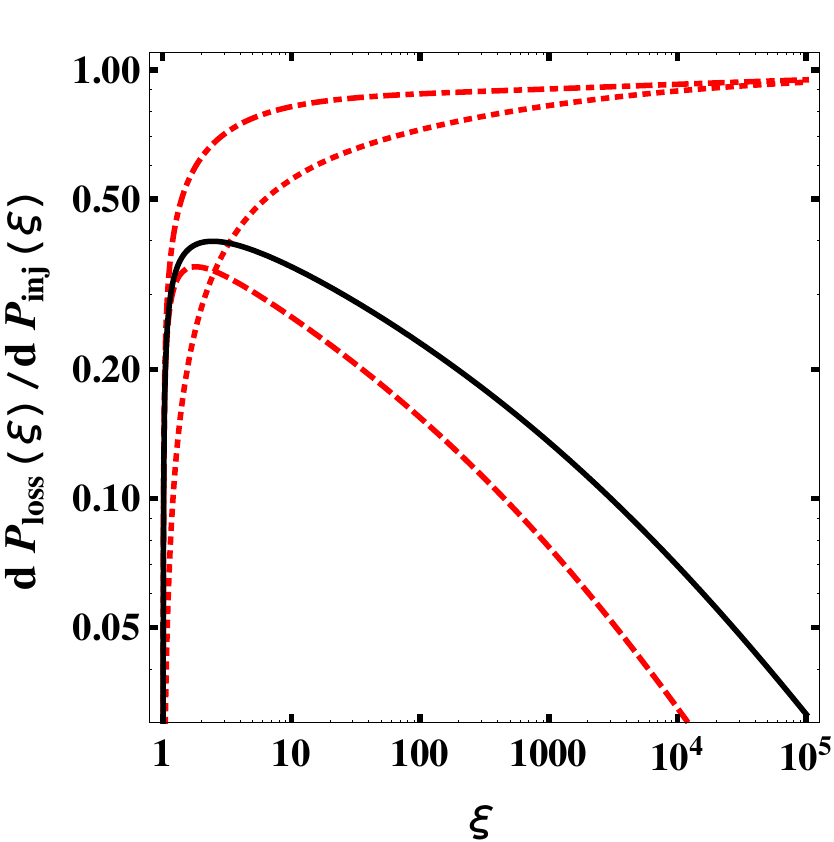}}
\caption{The local efficiency of the synchrotron and adiabatic processes vs.\ the length along the jet, $\xi$. The black solid curve is for the efficiency of the synchrotron emission for our solution with advection and without adiabatic losses. The dashed and dotted red curves are for the synchrotron and adiabatic losses, respectively, for the solution with both advection and adiabatic losses, and the dot-dashed red curve gives the total efficiency for that case. See Section \ref{no_adiabatic} for details and the parameters.
} \label{ratio}
\end{figure}

We then consider the local radiative efficiency defined as the ratio of ${\rm d}P_{\rm loss}/{\rm d}\xi$ to ${\rm d}P_{\rm inj}/{\rm d}\xi$, i.e., ${\rm d}P_{\rm loss}(\xi)/{\rm d}P_{\rm inj}(\xi)$. We show it by the black solid curve in Fig.\ \ref{ratio} for $M= 2\times 10^8\msun$, $z_0=10 r_{\rm g}$ (where $r_{\rm g}\equiv G M/c^2$ and $M$ is the BH mass), $\Gamma=30$, $B_0=90$ G, $p=2$, $\gamma_{\rm min}=1$, $\gamma_{\rm max0}\simeq 1.6\times 10^4$. For these parameters, $q\simeq 0.0034$, implying $\gamma_{\rm adv}\simeq 290$ for $\xi\gg 1$. We see that the jet radiative efficiency in this case decreases fast with the distance, and a large fraction of the electron internal energy injected to the jet is carried up to $z_{\rm max}$ and beyond it. For $z_{\rm max}=10^5 z_0$ ($\simeq 10$ pc) and the above parameters, the global efficiency of the synchrotron emission is $P_{\rm loss}/P_{\rm inj}\simeq 0.18$.

\begin{figure}
\centerline{\includegraphics[width=\columnwidth]{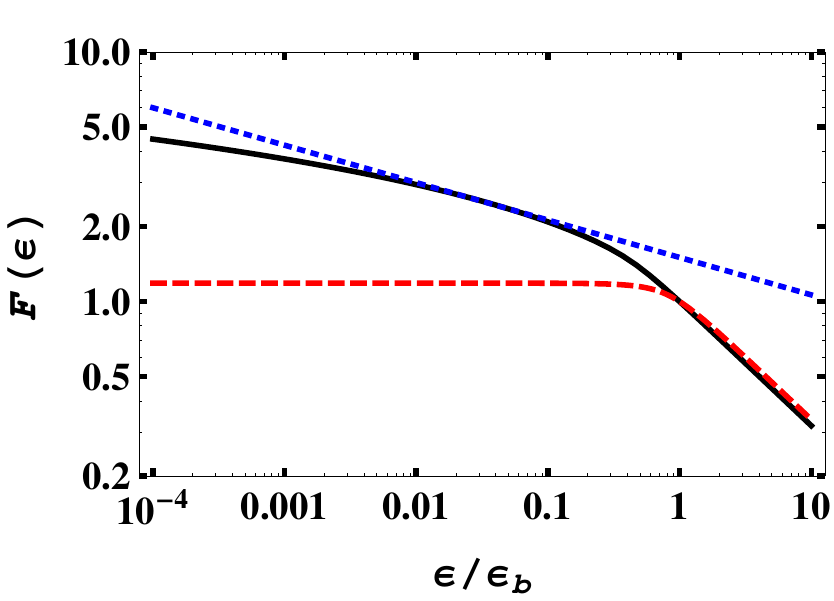}}
\caption{A comparison of the spectrum for the case without adiabatic losses and including advection along a jet (solid black curve) with the corresponding model with a maintained electron distribution (dashed red curve), for $p=2$ and $\xi_{\rm max}\rightarrow \infty$. The two spectra are normalized to unity at $\epsilon=\epsilon_{\rm b}$. The blue dotted curve shows a power law $\propto E^{-0.15}$.
} \label{spec_compare}
\end{figure}

We can compare this solution with that with advection and adiabatic losses. The solution for $k=1$, $p=2$ is given by equation 42 of Z14. The efficiency of the synchrotron and adiabatic processes are shown for the same parameters as above in Fig.\ \ref{ratio}. We see that the synchrotron emission is now lower than that in the model without advection. The total efficiency is close to unity, and that model is relatively close to that with adiabatic losses and complete local cooling (i.e., no effect of advection, equation \ref{cooling}), see, e.g., equation 45 of Z14. 

However, the solutions above assume that the synchrotron emission is not self-absorbed, while the flat radio spectra of accreting BHs are partially self-absorbed. In order to calculate the optical depth to self-absorption for a given electron energy, we use a mono-energetic approximation to the synchrotron emission and absorption, 
\begin{equation}
\epsilon\simeq (B/B_{\rm cr})\gamma^2,\quad \epsilon\equiv E(1+z_{\rm r})/\delta m_{\rm e}c^2,
\end{equation}
where $\epsilon$ is the dimensionless photon energy in the jet frame, $\delta$ is the Doppler factor, $z_{\rm r}$ is the redshift, $B_{\rm cr}$ is the critical magnetic field, and $E$ is the observed dimensional photon energy. We also need to know the electron distribution in physical units. From equations (\ref{N}) and (\ref{Pinj12}), we find
\begin{equation}
N(\gamma,\xi)=\frac{P_{\rm inj}}{2\upi \Gamma^2\beta_{\rm j} m_{\rm e}c^3 z_0^2 I_{\rm inj}\tan^2\Theta} \frac{\tilde{N}(\gamma,\xi)}{\tilde{Q}_0\xi^2}.
\label{N_dim}
\end{equation}
Then, the optical depth direction passing through the jet centre follows from equation (\ref{tausyn}),
\begin{equation}
\tau_{\rm sa}(\gamma,\xi)\simeq \frac{C_2(p)\sigma_{\rm T} B_{\rm cr} P_{\rm inj}}{2\alpha_{\rm f} m_{\rm e}c^3\Gamma^2\beta_{\rm j} z_0 B_0 I_{\rm inj} \gamma^4\delta\sin i\tan\Theta}\frac{\tilde{N}(\gamma,\xi)}{\tilde{Q}_0},
\label{taus}
\end{equation}
where $i$ is the inclination in the observer's frame, and $C_2(p)\sim 1$ is given, e.g., in \citet{zls12}. For example, at $P_{\rm inj}=5\times 10^{43}$ erg s$^{-1}$, $p=2$, $\gamma_{\rm sa}< \gamma_{\rm adv}$, $B_0=90$ G, $\Theta=0.2/\Gamma$, $\sin i=1/\delta$, $\tau_{\rm sa}=1$ occurs for $\gamma_{\rm sa}\simeq 70\ln^{1/6}\xi$, i.e., it is almost independent of $\xi$, as well as $\gamma_{\rm sa}< \gamma_{\rm adv} \simeq 290$. Given the very fast increase of $\tau_{\rm sa}$ with decreasing $\gamma$, the optical depth for $\gamma<\gamma_{\rm adv}$ quickly becomes very large. Then, the effect of self-absorption strongly reduces the synchrotron loss rate \citep{katarzynski06}. For the pure synchrotron model, this implies that $\dot\gamma\simeq 0$ in that regime, and the solution to equation (\ref{kinetic}) becomes that of equation (\ref{advection}),
\begin{equation}
\tilde{N}(\gamma,\xi)= \frac{\tilde{Q}_0}{\gamma^p}\ln\xi.
\label{Nsa}
\end{equation}
This solution turns out to be identical to that for $p=2$ and $\gamma<\gamma_{\rm adv}$, see equation (\ref{N2}), assumed in deriving equation (\ref{Nsa}). This coincidence is due to the middle term in equation (\ref{kinetic}) being null for either $\tilde{N}(\gamma)\propto \gamma^{-2}$ and $\dot\gamma=0$.

We can then calculate the synchrotron spectrum corresponding to a given electron distribution. We use equation (\ref{1D}) and consider a conical, constant-speed, jet. This yields 
\begin{align}
&F_{\rm S}(E)=\left(m_{\rm e} c\over h\right)^3 {\upi (1+z_{\rm r})c C_1 z_0^2 \delta^3 B_{\rm cr}^{1/2} \epsilon^{5/2}\tan\Theta\sin i\over 6 C_2 B_0^{1/2} D_L^2} \times\nonumber\\
&\qquad \int_{\xi_0(\epsilon)}^{\infty} {\rm d}\xi\,\xi^{3/2}
\left\{1-\exp\left[-\tau_{\rm sa}(\epsilon,\xi)\right]\right\}, \label{logN}
\end{align}
where $h$ is the Planck constant, $C_1\sim 1$ is a normalization factor of the synchrotron emissivity, which is a weak function of the average electron index of $N$ (see, e.g., \citealt{zls12}), $D_L$ is the luminosity distance, 
\begin{equation}
\xi_0(\epsilon)=\max\left(1, \frac{\gamma_0^2 B_0}{\epsilon B_{\rm cr}}\right),
\end{equation}
see equation (\ref{ximin}), and $\gamma_0$ is the minimum Lorentz factor for which the electron distribution is calculated; we assume $\gamma_0=1$. In order to account for the high-energy cutoff, we multiply the above spectrum for $\xi\leq \xi_{\rm max}$ by $\exp(-\epsilon/\epsilon_{\rm max})$, where $\epsilon_{\rm max}$ is the distance-independent cutoff jet-frame photon energy resulting from the acceleration with $\gamma_{\rm max}(\xi)$ given by equation (\ref{gemax}),
\begin{equation}
\epsilon_{\rm max}=\frac{9\eta_{\rm acc}}{4\alpha_{\rm f}}.
\end{equation}
On the other hand, our solution for the electron distribution at $z>z_{\rm max}$, equation (\ref{Nhigh}), has an explicit dependence of $\gamma_{\rm max}(\xi)$, above which $N(\gamma)=0$, and thus we use equation (\ref{logN}) directly. 

The optical depth for the electron distribution (\ref{Nsa}) can be expressed through the photon energy at which the entire jet becomes optically thin; here, for simplicity, we define this energy, $\epsilon_{\rm b}$, for the case without the $\ln\xi$ factor in equation (\ref{Nsa}),
\begin{equation}
\epsilon_{\rm b}=\left(\frac{C_2 \sigma_{\rm T} P_{\rm inj}}{2\alpha_{\rm f} m_{\rm e}c^3\Gamma^2\beta_{\rm j}z_0 I_{\rm inj} \delta \sin i\tan\Theta} \right)^{\frac{2}{4+p}}\left(\frac{B_0}{B_{\rm cr}}\right)^{\frac{2+p}{4+p}}.
\label{epsilon_b_sa}
\end{equation}
Note that this is just a parameter and that the jet in this solution becomes optically thin at all values of $\xi\geq 1$ already at $\epsilon\ga 0.5\epsilon_{\rm b}$. Furthermore, there is a small range of $\tau_{\rm sa}<1$ close to $\xi=1$ also for values of $\epsilon< 0.5\epsilon_{\rm b}$ (due to the presence of the $\ln\xi$ term in $\tau_{\rm sa}$). From the above definition, we obtain,
\begin{equation}
\tau_{\rm sa}(\epsilon,\xi)=(\epsilon\xi/\epsilon_{\rm b})^{-(p+4)/2} \ln\xi,\quad \xi\leq \xi_{\rm max}.
\label{taulog}
\end{equation}
The model without the $\ln \xi$ factor in equation (\ref{taulog}) is equivalent to that of BK79, which yields $\alpha=0$ in the partially self-absorbed regime, $\epsilon\ll \epsilon_{\rm b}$. We find the spectrum of equations (\ref{logN}--\ref{taulog}) is softer, as shown in Fig.\ \ref{spec_compare}. It can be roughly approximated by a power law with $\alpha\simeq -0.15$.

We then use the full electron distribution for $p=2$, of equation (\ref{N2}), and calculate the resulting spectrum. This is given by equation (\ref{logN}) with $\tau_{\rm sa}$ given by
\begin{align}
&\tau_{\rm sa}(\epsilon,\xi) =\left(\frac{\epsilon\xi}{\epsilon_{\rm b}}\right)^{-(p+4)/2}
\begin{cases}
\ln\xi,&\epsilon\leq \epsilon_{\rm adv}(\xi);\cr
\ln\left[1+\displaystyle{\frac{(\xi B_0)^{1/2}}{q (\epsilon B_{\rm cr})^{1/2}}}\right],
&\epsilon> \epsilon_{\rm adv}(\xi),\cr
\end{cases}\label{tauN2}\\
&\epsilon_{\rm adv}(\xi)\equiv \frac{B_0}{B_{\rm cr}} \frac{\gamma_{\rm adv}^2}{\xi},
\end{align}
where $\gamma_{\rm adv}$ is given in equation (\ref{gadv}). We have kept above the power-law dependence with the index $p$, which gives an approximate solution for $p\simeq 2$. 

\section{A general solution with adiabatic losses}
\label{adiabatic}

Z14, based on the results of \citet{stawarz08}, have obtained the solution for the case of a conical jet with conserved magnetic energy flux, synchrotron and adiabatic losses and advection for $k=1$, see their equation 40. The dimensionless rate of the adiabatic energy loss is
\begin{equation}
\tilde{\gamma}_{\rm a}(\gamma,\xi)=- \frac{2\gamma}{3\xi}.
\label{gamma_dot_a}
\end{equation}
Here, we obtain the solution for a general value of $k$. For $\gamma_{\rm max}\rightarrow \infty$, we have,
\begin{equation}
\tilde{N}(\gamma,\xi)=
\tilde{Q}_0\gamma^{-2}\xi^{-2/3} \int_{\xi_{\rm m}}^\xi {\rm d}\xi' {\xi'}^{2/3-k} \gamma'(\gamma,\xi;\xi')^{2-p},\label{N_analytical_a}
\end{equation}
which yields
\begin{align}
&\tilde{N}(\gamma,\xi)=
3\tilde{Q}_0\gamma^{-p}\xi^{(4-2 p)/3}\times\label{Nadv}\\
&\left. \frac{\xi'^{(1-3k+2 p)/3} 
{_2F_1}\left[\frac{3 k-2 p-1}{5},2-p,\frac{3 k-2 p+4}{5}; 
\frac{3 q\gamma}{5\xi+3 q\gamma}\left(\frac{\xi}{\xi'}\right)^{5/3}
\right]}{(1+2 p- 3 k)\left(1+\frac{3 q\gamma} {5\xi}\right)^{2-p}}\right|^{\xi'=\xi}_{\xi'=\xi_{\rm m}},\nonumber
\end{align}
where now the formulae for $\gamma'$, $\xi_{\rm m}$ and $\gamma_{\rm adv}$ are now different than those in Section \ref{no_adiabatic}, see equations 36 and 37 of Z14,  
\begin{align}
&\gamma'(\gamma,\xi;\xi')=\frac{\gamma(\xi/\xi')^{2/3}}{1-\frac{3 q\gamma}{5\xi}\left[(\xi/\xi')^{5/3}-1\right]},
\label{gamma_prime_a}\\
&\xi_{\rm m}(\xi)= \begin{cases}1, &\gamma\leq \gamma_{\rm adv}(\xi);\cr
\xi\left(1+\frac{5\xi}{3 q\gamma}\right)^{-3/5}, &\gamma> \gamma_{\rm adv}(\xi),\cr
\end{cases}\label{xima}\\
&\gamma_{\rm adv}(\xi)\equiv \frac{5\xi}{3 q(\xi^{5/3}-1)}.
\label{gadva}
\end{align}

Of interest for our comparison with the data (see Section \ref{comparison}) is the case of $k=3/4$, $p=2$. We find
\begin{equation}
\tilde{N}(\gamma,\xi)=\frac{12\tilde{Q}_0\xi^{1/4}}{11\gamma^2} \begin{cases}
1-\xi^{-11/12}, &\gamma\leq \gamma_{\rm adv}(\xi);\cr
1-\left(1+\frac{5\xi}{3 q\gamma}\right)^{-11/20}, &\gamma> \gamma_{\rm adv}(\xi);\cr
\frac{11\xi}{12 q\gamma}, &\gamma\gg \xi/q,\cr
\end{cases}
\label{N2a}
\end{equation}
where the last equality is approximate.

\section{Comparison with spectral data}
\label{comparison}

\begin{figure}
\centerline{\includegraphics[width=\columnwidth]{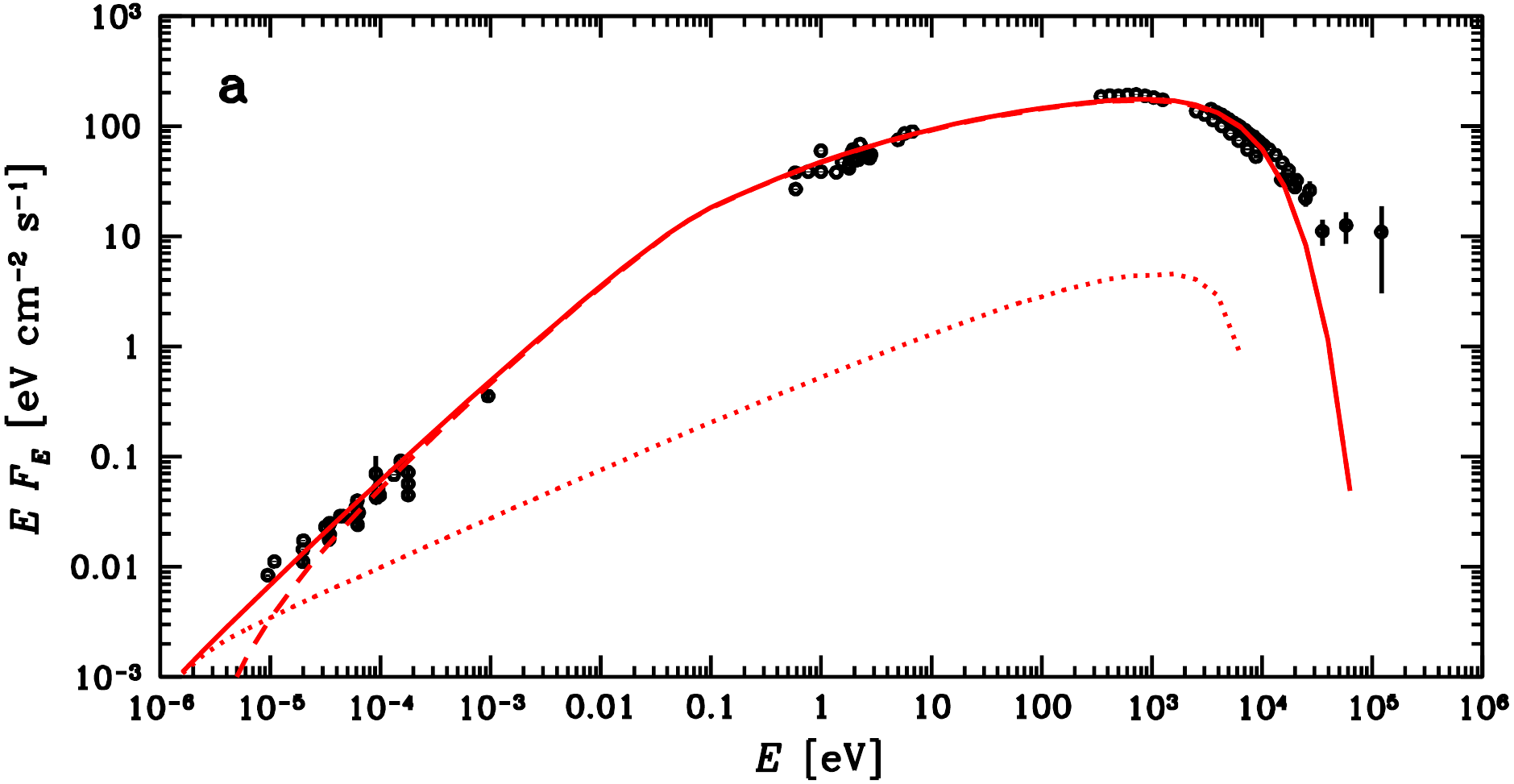}}
\centerline{\includegraphics[width=\columnwidth]{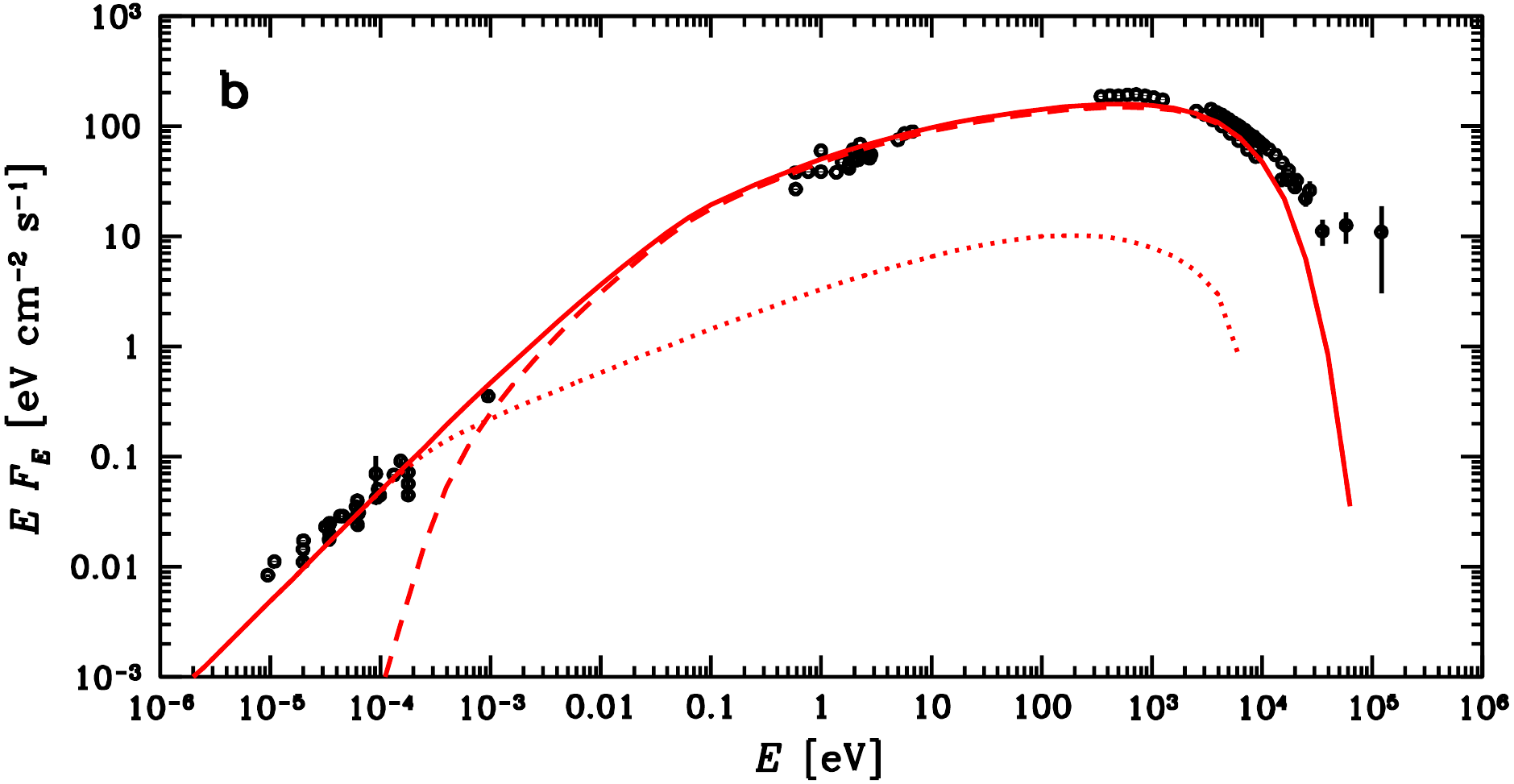}}
\centerline{\includegraphics[width=\columnwidth]{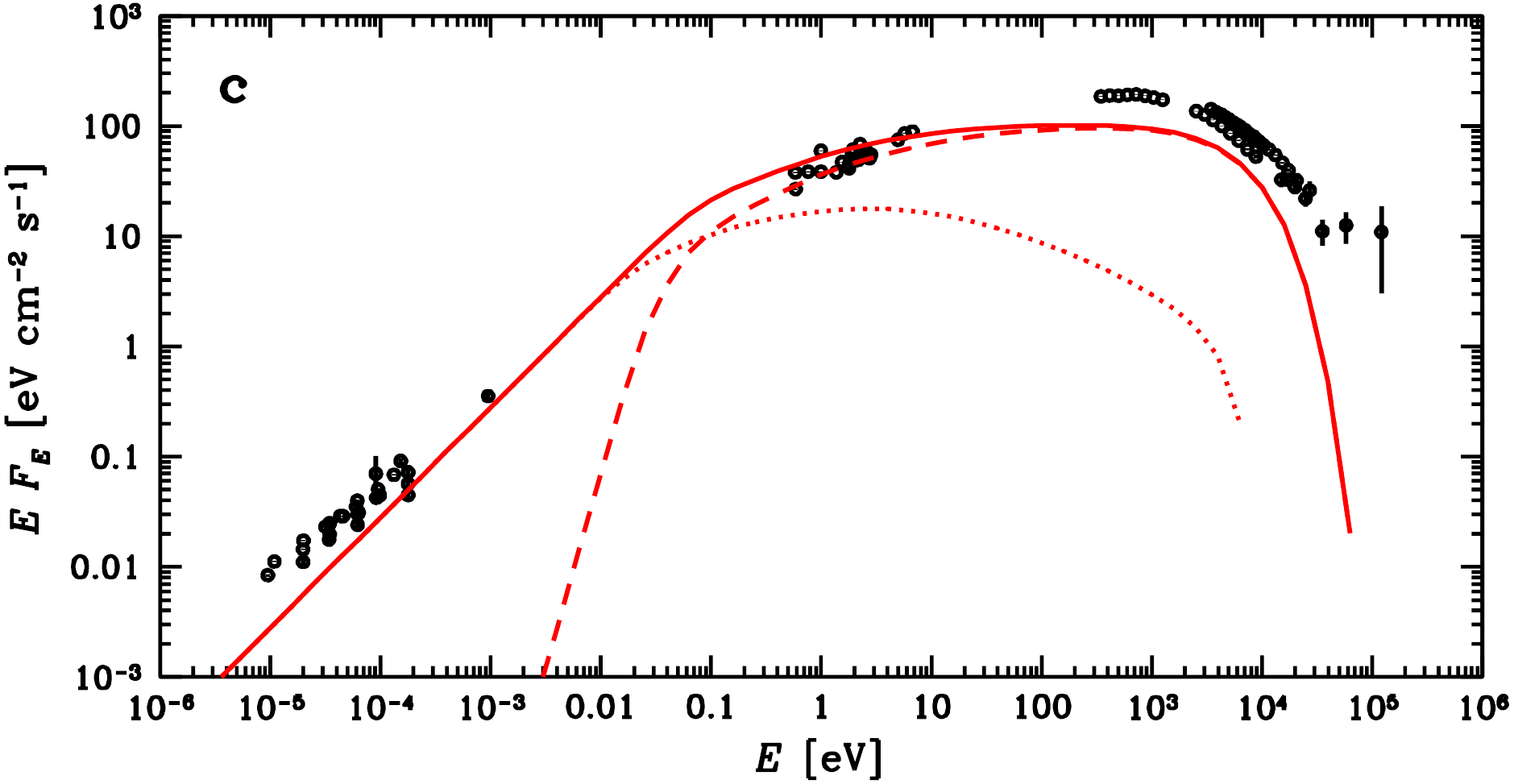}}
\caption{The radio-to-X-ray spectrum of the quiescent state of Mrk 421 compared to the predictions of the model with the adiabatic losses balanced by reacceleration. The distance up to which fresh electrons are accelerated and injected in the jet, $z_{\rm max}$, is (a) 10 pc; (b) 0.1 pc; and (c) 0.001 pc, respectively. The power provided to the electrons, $P_{\rm inj}$, is chosen in such way that the observed flux at 1 eV is approximately the same in all three cases. The dashed and dotted curves show the jet emission corresponding to $\leq z_{\rm max}$ and $>z_{\rm max}$, respectively, and the solid curve shows the sum. See Section \ref{comparison} for other parameters. 
}
\label{models}
\end{figure}

\begin{figure}
\centerline{\includegraphics[width=\columnwidth]{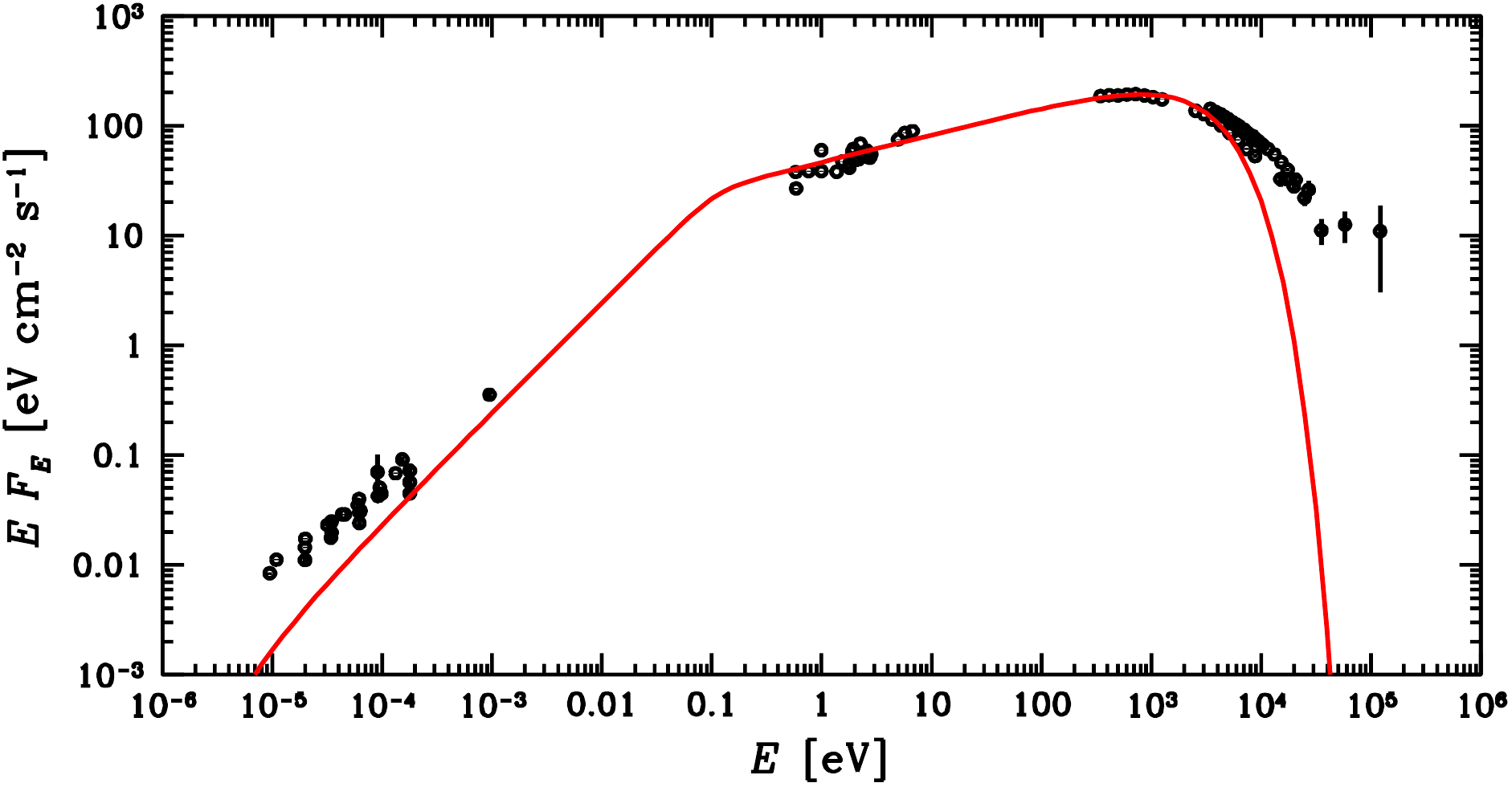}}
\caption{The radio-to-X-ray spectrum of the quiescent state of Mrk 421 \citep{abdo11b} compared to the synchrotron model of (BK79), in which the electron distribution is maintained along a conical jet with constant velocity and conserved energy flux of toroidal magnetic field. We see that this model fails to reproduce the radio--mm part of the spectrum.
See Section \ref{comparison} for the parameters. 
}
\label{f:bk79}
\end{figure}

\begin{figure}
\centerline{\includegraphics[width=\columnwidth]{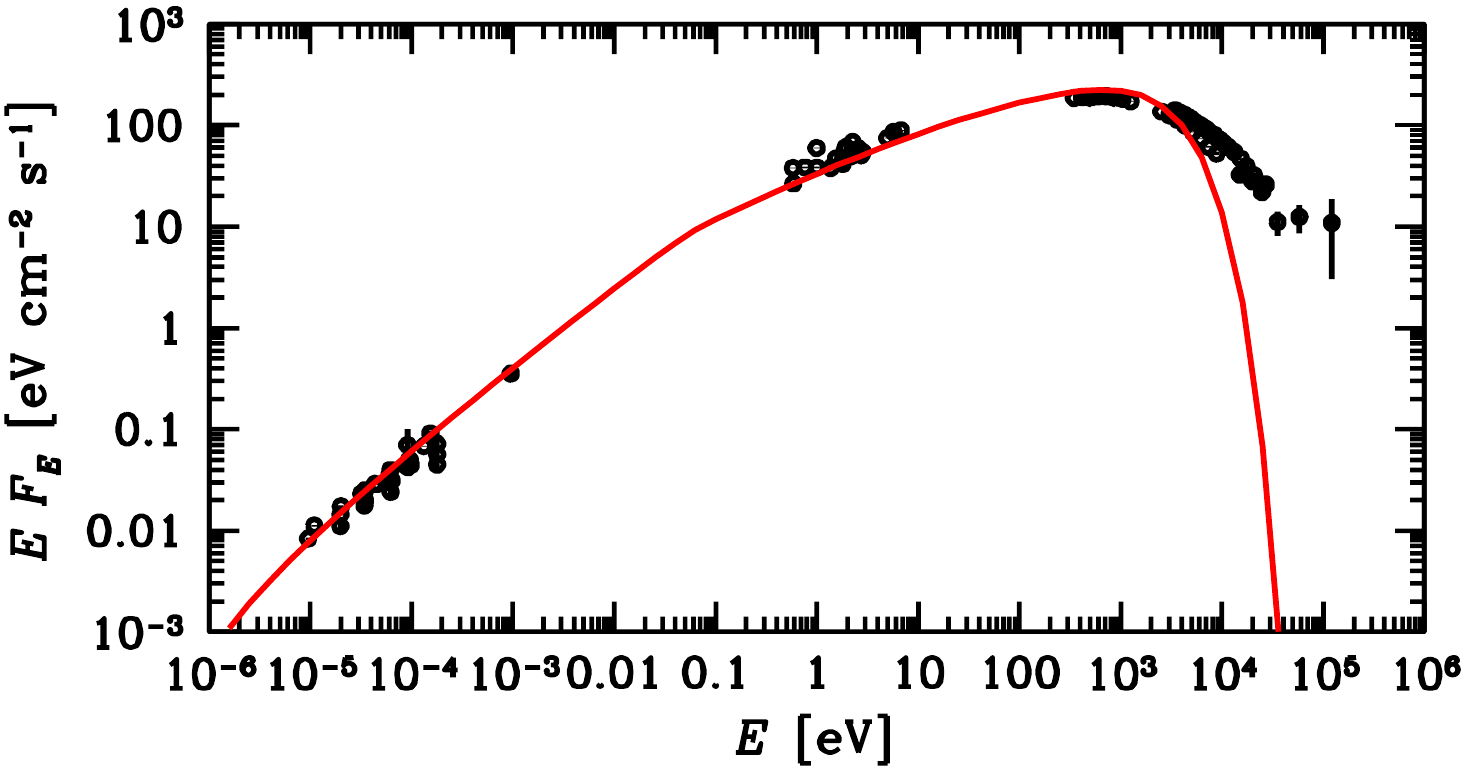}}
\caption{The radio-to-X-ray spectrum of the quiescent state of Mrk 421 \citep{abdo11b} compared to the synchrotron model including adiabatic losses and advection along the jet. This model reproduces the observed spectrum but it requires a very large power supplied to the relativistic electrons in the jet, see Section \ref{comparison}. 
}
\label{spectrum_ad}
\end{figure}

We then compare our model synchrotron spectrum with that of the quiescent spectrum of a blazar. We choose Mrk 421, which is a high-synchrotron peaked BL Lac object at the redshift of $z=0.0308$. We have chosen this object because it has a soft radio-to-IR spectrum with $\alpha\approx -0.2$, as well as it is one of the brightest sources in its class, its continuum from radio to hard X-rays is well covered, and the high-quality data we use have no measureable contribution from either the accretion disc or the host galaxy \citep{abdo11b}. Furthermore, all of those data have been obtained and averaged over a single observational campaign lasting about 1.5 yr, during which the measured fluxes were almost constant \citep{abdo11b}. We use the luminosity distance of $D_L=133$ Mpc, corresponding to $H_0=71$ km s$^{-1}$ Mpc$^{-1}$ and $\Omega_\Lambda =0.73$. The BH mass was measured via the $M$-$\sigma$ relation by \citet*{barth03} as $\log_{10} (M/\msun)=8.28\pm 0.11$; here we assume $M=2\times 10^8\msun$, corresponding to $r_{\rm g}\simeq 3\times 10^{13}$ cm. We assume the jet is viewed at $i=1/\Gamma$, the jet half-opening angle is $\Theta=0.2/\Gamma$ (following \citealt{pushkarev09,clausen13}), and, for simplicity, we assume the minimum Lorentz factor of the accelerated electrons to be $\gamma_{\rm min}=1$.

We find we can fit the observations rather well by our model, as given by equations (\ref{taulog}) and (\ref{tauN2}). We show an example of a good fit in Fig.\ \ref{models}(a). The parameters are $P_{\rm inj}=5\times 10^{43}$ erg s$^{-1}$, $p=2.0$, $z_0= 10 r_{\rm g}$, $z_{\rm max}=10$ pc ($\simeq 10^5 z_0$ or $10^6 r_{\rm g}$), $B_0=90$ G, $\Gamma=30$, $\Theta=0.2/\Gamma$, $\eta_{\rm acc}=1.6\times 10^{-6}$. The chosen values of $\Gamma$, $z_0$ and $B_0$ correspond to the advection factor of $a\approx 0.0034$. We use the above values of the parameters for numerical examples in Section \ref{no_adiabatic}, in particular, in Figs.\ \ref{p2_3}(a) and \ref{ratio}. We show separately the emission below and above $z_{\rm max}$; the radiative efficiency of those emission is $P_{\rm loss}/P_{\rm inj} \approx 0.18$ and $\approx 0.003$, respectively. Thus, the total synchrotron luminosity radiated in all directions by the jet is $P_{\rm loss}\simeq 10^{43}$ erg s$^{-1}$. The remaining injected power is retained by the accelerated electrons beyond $z_{\rm max}$, with $P_{\rm e}\simeq 4\times 10^{43}$ erg s$^{-1}$, and then it may be lost in interactions with the surrounding medium, in particular in radio lobes. The jet in this model is weakly magnetized, $P_B\simeq 4\times 10^{41}$ erg s$^{-1}$. The kinetic power in ions (equation \ref{P_i}) depends on the unknown pair content and is $\simprop \gamma_{\rm min}^{-1}$ (on which the injected power depends only weakly). It can be thus comparable, lower or larger than $P_{\rm e}$.

We find the value of $z_{\rm max}$ chosen above to be approximately the minimum one compatible with the observed radio--mm spectra. Since the jet emission at large distances becomes inefficient, any larger value of $z_{\rm max}$ would result in almost no change of the spectrum, and would thus also provide a good fit to the data. On the other hand, reducing $z_{\rm max}$ leads to spectra that are too hard in the radio range, as shown in Figs.\ \ref{models}(b) and (c), for $z_{\rm max}=0.1$ and 0.001 pc, and $P_{\rm inj}=3\times 10^{43}$ and $1.5\times 10^{43}$ erg s$^{-1}$, respectively (with $P_{\rm inj}$ chosen to yield approximately the same observed flux at 1 eV in all three cases). The effect of the hardening of the radio spectra is due to the inefficient self-absorbed synchrotron emission above $z_{\rm max}$, in the cases of the acceleration of fresh electrons terminating too close to the BH. In the case of the spectrum shown in Fig.\ \ref{models}(c), we could have adjusted the parameters to yield a good fit to the spectra in the X-ray range. However, we could not obtain a soft enough radio--mm spectrum since it is produced by the coasting (above $z_{\rm max}$) part of the jet, in which case the self-absorbed part of the synchrotron spectrum is $\alpha\approx 0$, as in the original model of BK79.

Also, the relatively soft radio-mm spectrum of Mrk 421 cannot be fitted with the model of BK79 (which we describe in Appendix \ref{approx}). We show it in Fig.\ \ref{f:bk79}, which assumes a conserved electron distribution along the jet and $B=B_0\xi^{-1}$ (i.e., $a=2$, $b=1$ in the notation of Appendix), which yield the energy index in the partially self-absorbed part of the synchrotron spectrum of $\alpha=0$. The parameters of the shown example model are $p=2.5$, $z_0=20 r_{\rm g}$, $B_0=40$ G, $\Gamma=30$, $z_{\rm max}=10$ pc, $\eta_{\rm acc}=7\times 10^{-7}$. We see the model low-energy spectrum is significantly too hard to account for the radio-mm observations. Changing the parameters of this model would not help since an inherent feature of this model is $\alpha=0$ in the partially synchrotron self-absorbed part of the spectrum.

We then consider the model including adiabatic losses (as well as advection and synchrotron losses, as described in Section \ref{adiabatic}. Models with $k=1$ yield radio-mm spectra with $\alpha=0$, identical to the model of BK79, e.g., one shown in Fig.\ \ref{f:bk79}. In order to fit the soft radio-mm spectrum, we need to inject relativistic electrons at a rate $\propto \xi^{-k}$ with $k=0.75$. This means that the power supplied to the electrons per a logarithmic distance interval increases with the distance as $\xi^{0.25}$. An example of a fit with this model is shown in Fig.\ \ref{spectrum_ad}. The parameters are $P_{\rm inj}=5.3\times 10^{44}$ erg s$^{-1}$, $p=2.05$, $z_0= 10 r_{\rm g}$, $z_{\rm max}=100$ pc ($\simeq 10^6 z_0$ or $10^7 r_{\rm g}$), $B_0=80$ G, $\Gamma=30$, $\Theta=0.2/\Gamma$, $\eta_{\rm acc}=6\times 10^{-7}$. The chosen values of $\Gamma$,  $z_0$ and $B_0$ correspond to the advection factor of $q\approx 0.0027$. The radiative efficiency of the synchrotron emission is $P_{\rm loss}/P_{\rm inj} \approx 0.018$, implying the total synchrotron luminosity approximately the same as in the model without adiabatic losses, namely $\simeq 10^{43}$ erg s$^{-1}$. On the other hand, the power needed to be supplied to the jet with adiabatic losses is about 10 times higher than that without it. Almost all of this power is then lost adiabatically. We also note that the chosen value of $z_{\rm max}$ is approximately the lowest allowed by the data. For $z_{\rm max}=10$ pc, the model spectrum goes substantially below the data at the lowest measured radio frequencies, similarly to Figs.\ \ref{models}(b, c).

\section{Discussion and conclusions}
\label{discussion}

We have investigated effects of adiabatic losses and electron advection on synchrotron spectra of jets, with emphasis on the partially synchrotron self-absorbed regime. Our approach to the problem has been to use a kinetic equation for the steady-state electron distribution. Our study has been motivated by stringent energetic requirements imposed in modelling of jets by the presence of adiabatic losses, especially in the case of relatively soft partially self-absorbed synchrotron spectra, with $\alpha<0$, in which case the local power required to be provided to the relativistic electrons in order to compensate for their adiabatic losses has to increase with the distance. This results in the total supplied power greatly exceeding the power in radiation. This difficulty can be avoided if the power dissipated adiabatically is used for reacceleration of the relativistic electrons (in addition to the main acceleration process). We also note that spectra with $\alpha<0$ cannot be caused by dissipation of the magnetic energy along the jet.

We have then obtained analytical solutions of the jet kinetic equation in the case of the acceleration rate of a power-law form in both the distance and the Lorentz factor. These solutions are new in the case with adiabatic losses balanced by reacceleration. In the case with adiabatic losses, we have given a generalization of the solution of Z14. We have related the obtained steady-state electron distributions to the power supplied in the acceleration and to the components of the jet kinetic power in relativistic electrons and in cold ions. We have taken into account the effect of self-absorption on the obtained distributions but we have found the effect on the electron distribution is minor, due to the dominance of advection at low values of the Lorentz factor. We have then obtained a simple formula for the observed synchrotron spectrum based on radiation transfer. It is in the form of a single integral over the jet length and it is valid in both the partially optically thick and optically thin regimes.

We have also shown that the absence of adiabatic losses leads to accumulation upstream of relativistic electrons accelerated downstream. This, in turn, leads to an increase with the distance of the electron number integrated over the jet cross section. This effect is similar to that obtained with the rate of acceleration per decade increasing with the jet length in the case with adiabatic losses, but with much lower energetic requirements. Such an increase is required in order to explain spectra with $\alpha<0$.

We have applied our results to the quiescent radio to X-ray spectrum of the BL Lac Mrk 421, which has the radio-to-IR spectrum with $\alpha \approx -0.2$. We have found that we can reproduce the observed spectrum with a model without adiabatic losses and with a constant rate of the electron acceleration per decade, which is possible due to the electron accumulation process described above. The radiative efficiency of this model is $\sim$0.2, i.e., that part of the energy supplied to the electrons is radiated (with an estimated luminosity of $\sim 10^{43}$ erg s$^{-1}$) in the synchrotron process. The power that needs to be dissipated in the jet is then relatively modest, and it can be provided by, e.g., internal shocks or magnetic reconnection. Still, the relativistic electrons retain most of their internal energy up to an arbitrary distance. Thus, a prediction of this model is that such jets contain substantial internal energy (in comparison to the kinetic energy in cold ions) when hitting the radio lobes. This agrees with the results of \citet{snios18} for Cygnus A.

We have determined the minimum length of the part of the jet over which the acceleration takes place for Mrk 421, obtaining $\sim$10 pc. A larger length would also fit the data, as a consequence of the low synchrotron radiative efficiency at large distances. On the other hand, a lower length would lead to radio-to-mm spectra harder than the observed one with $\alpha\approx -0.2$. Thus, our results show that reacceleration (at the rate equal to the adiabatic loss rate) of electrons initially injected at the jet base is not sufficient to explain the observed spectra. Additional acceleration along the jet is required (as assumed in this work). 

We have then reproduced the observed spectrum with a model with adiabatic losses. However, that case required the power per length decade supplied to the accelerated electrons to increase with distance, approximately as $z^{0.25}$. Almost all of the supplied power is then lost adiabatically, with the global radiative efficiency being only $\sim$0.02, and the minimum power needed to be supplied to the relativistic electrons becomes as large as $\sim 5\times 10^{44}$ erg s$^{-1}$.

In this work, we have limited our study to the synchrotron process. Additional constraints can be provided by considering the synchrotron self-Compton process and requiring that it reproduces observed quiescent $\gamma$-ray spectra, which issue we will consider in a subsequent study.

Finally, we point out the simplifications of our approach. We have neglected the diffusion terms in the electron kinetic equation. However, given the usual high jet velocity of blazars, $\beta_{\rm j}\approx 1$, advection is expected to dominate over diffusion. We have considered jets with homogeneous cross sections, while both clumping and spine/sheath structures are possible and observed in some blazars (e.g., \citealt{tavecchio08}). 

We have also assumed the jet to have a constant velocity. Astrophysical jets can be both accelerated and decelerated; in particular, a fraction of the adiabatic losses can be used to accelerate the jet. A study of the associated effects is beyond the scope of our paper. Still, we point out that jet acceleration can give an alternative mechanism to produce soft, partially self-absorbed synchrotron spectra, provided the direction to the observer remains within the emission beam, $\sim\! 1/\Gamma$. Then, a $\Gamma$ increasing with the distance will lead to a corresponding increase of the observed flux, due to relativistic beaming, which will result in spectra with $\alpha<0$. Alternatively, a decelerating jet would give $\alpha>0$. These effects would change if the line of sight moves out or in the emission beam.

\section*{ACKNOWLEDGEMENTS}

We thank Rafa{\l} Moderski and Micha{\l} Ostrowski for valuable discussions. This research has been supported in part by the Polish National Science Centre grants 2013/10/M/ST9/00729 and 2015/18/A/ST9/00746.

\appendix

\section{The synchrotron flux}
\label{approx}

In order to obtain the jet synchrotron flux including both the optically-thick and optically-thin parts, we can use the standard radiative transfer equation, e.g., \citet{rybicki79}, and integrate it over the projected jet area. This was done in \citet{zls12} for the model of BK79 and in Z14 for an arbitrary electron distribution and a conical jet with constant speed, and neglecting the redshift in both works. Here, we generalize those results for a jet with arbitrary dependences along its length of the jet radius, $r(\xi)$, the bulk Lorentz factor, $\Gamma$, and the jet-frame magnetic field and the electron distribution, $B(\xi)$ and $N(\gamma,\xi)$, respectively. In this case, the relationship between the observed photon energy, $E$, and the dimensionless one in the jet frame, is a function of the position, 
\begin{equation}
\epsilon(\xi)\equiv \frac{E(1+z_{\rm r})}{\delta(\xi) m_{\rm e}c^2},\quad \delta(\xi)\equiv \frac{1}{\Gamma(\xi)(1-\beta_{\rm j}(\xi)\cos i)}.
\label{epsilon}
\end{equation}
With that definition of $\epsilon$, the observed flux is given by,
\begin{align}
&F_{\rm S}(E)=\left(m_{\rm e} c\over h\right)^3 {(1+z_{\rm r})c C_1 z_0 (\epsilon\delta)^{5/2}\sin i\over 3 C_2 D_L^2} \times\nonumber\\
&\quad \int_{\xi_0(\epsilon)}^{\xi_{\rm max}} {\rm d}\xi\,r(\xi)\left[\delta(\xi)B_{\rm cr}\over B'(\xi)\right]^{1/2}\!\! \int_{-1}^1 {\rm d}\psi \left\{1-\exp\left[-\tau_{\rm sa}(\epsilon,\xi,\psi)\right]\right\},
\label{thin_thick}
\end{align}
where $\epsilon\delta$ is the $\xi$-independent photon energy in the BH frame, $\psi\equiv x/r(\xi)$, $x$ is the dimension perpendicular to both the jet axis and the line of sight, $z_{\rm r}$ is the redshift, $i$ is the jet inclination in the observer's frame, $D_L$ is the luminosity distance, and $\xi_0(\epsilon)$ is the minimum $\xi$ at which a given photon energy can be produced, which is the maximum of 1 and the solution of
\begin{equation}
\frac{\epsilon B_{\rm cr}}{B(\xi_0)}=\gamma_0^2.
\label{ximin}
\end{equation}
Then, the self-absorption optical depth is,
\begin{equation}
\tau_{\rm sa}(\epsilon,\xi,\psi)= \frac{C_2 \upi \sigma_{\rm T} r(\xi)B(\xi)(1-\psi^2)^{1/2}}{\alpha_{\rm f} B_{\rm cr}\epsilon(\xi)^2 \delta(\xi) \sin i} N\left(\sqrt{\epsilon(\xi) B_{\rm cr}\over B(\xi)},\xi\right).
\label{tausyn}
\end{equation}
We note that equation (\ref{thin_thick}) assumes that the energy unit in the energy flux per unit photon energy, $F_{\rm S}(E)\equiv {\rm d}F_{\rm S}/{\rm d}E$, is the same as the unit of the photon energy, $E$. E.g., both $F_{\rm S}$ and $E$ can be in erg or both in eV. If those energy units are different, then $F_{\rm S}$ needs to be multiplied by the ratio of the units, e.g., for $F_{\rm S}$ in erg cm$^{-2}$ s$^{-1}$ and $E$ in eV, equation (\ref{thin_thick}) needs to be multiplied by erg/eV. Also, equation (\ref{thin_thick}) is strictly valid only for $i\sim 90\degr$; however, \citet{z16} have shown it approximates well the exact solution for $i\gtrsim 1/\Gamma$. We also note that $h/m_{\rm e}c=\lambda_{\rm C}$, the electron Compton wavelength.

The integration over the projected area is double, over the jet length and radius. However, it can be replaced by an integral over the jet length only with little loss of accuracy, by including the normalization factor $\upi/2$, which corresponds to the integration in the optically-thin limit. In the fully optically-thick regime that integration yields a factor of 2. However, the partially self-absorbed part of the spectrum includes contributions from both optically-thick and optically-thin parts of the jet, and a typical accuracy of this approximation is better than 5 per cent. Including the factor of $\upi/2$ yields the synchrotron flux of
\begin{align}
&F_{\rm S}(E)=\left(m_{\rm e} c\over h\right)^3 {\upi (1+z_{\rm r})c C_1 z_0 (\epsilon\delta)^{5/2}\sin i\over 6 C_2 D_L^2} \times\nonumber\\
&\qquad \int_{\xi_0}^{\xi_{\rm max}} {\rm d}\xi\,r(\xi)\left[\delta(\xi)B_{\rm cr}\over B(\xi)\right]^{1/2} \left\{1-\exp\left[-\tau_{\rm sa}(\epsilon,\xi)\right]\right\},
\label{1D}
\end{align}
where 
$\tau_{\rm sa}(\epsilon,\xi)\equiv \tau_{\rm sa}(\epsilon,\xi,\psi=0)$.

Following \citet{konigl81}, we can then obtain approximate analytical solutions for the jet emission assuming power-law dependencies of the electron distribution and magnetic field strength on the distance along a conical jet with constant speed. Namely, we assume
\begin{equation}
r(\xi)= z_0\xi\tan\Theta,\quad N(\gamma,\xi)=N_0 \xi^{-a}\gamma^{-p},\quad B(\xi)=B_0 \xi^{-b},
\label{pl}
\end{equation}
where $a=2$, $b=1$ for the case of the electron number and the magnetic energy flux conserved along a conical, constant $\Gamma$, jet (BK79). In this case, $\epsilon$ is no longer a function of $\xi$. We can then express the optical depth for self-absorption for line of sights crossing the jet spine as
\begin{equation}
\tau_{\rm sa}(\epsilon,\xi)=(\epsilon/\epsilon_{\rm b})^{-(p+4)/2} \xi^{1-a-b(p+2)/2},
\label{tau}
\end{equation}
where $\epsilon_{\rm b}$ is the break energy defined by $\tau_{\rm sa}(\epsilon_{\rm b},1)=1$, i.e., at which the entire emitting part of the jet becomes optically thin. We find $\epsilon_{\rm b}$ as
\begin{equation}
\epsilon_{\rm b}=\left(\frac{C_2 \upi \sigma_{\rm T} z_0 N_0\tan\Theta}{\alpha_{\rm f} \delta \sin i}\right)^{\frac{2}{4+p}}\left(\frac{B_0}{B_{\rm cr}}\right)^{\frac{2+p}{4+p}},
\label{epsilon_b}
\end{equation}
which is independent of the values of $a$ and $b$.

For the case of the dependences of equation (\ref{pl}), we can integrate analytically equation (\ref{1D}), which yields the spectrum in terms of the lower incomplete gamma function, $\gamma_{\rm E}$,
\begin{align}
&F_{\rm S}(E)=\left(m_{\rm e} c\over h\right)^3 {\upi (1+z_{\rm r})c C_1(p) z_0^2\delta^3 B_{\rm cr}^{1/2}\epsilon_{\rm b}^{5/2}\tan\Theta\sin i\over 3 C_2(p)B_0^{1/2} D_L^2} \times\nonumber\\
& \left\{\frac{(\epsilon/\epsilon_{\rm b})^\alpha}{2-2a-b(p+2)}
\gamma_{\rm E}\left[\frac{4+b}{6-2 a-b(p+1)}, \left(\frac{\epsilon}{\epsilon_{\rm b}}\right)^{-(p+4)/2} \!\!\!\!\!\!\xi_0^{6-2 a-b(p+1)}\right]\right.+\nonumber\\
&\left. -\left(\frac{\epsilon}{\epsilon_{\rm b}}\right)^{5/2} \frac{\xi_0^{2+b/2}}{4+b}\right\}, \label{pl2}\\
&\alpha=\frac{5 a+3 b+2(b-1)p-13}{2a-2+b(p+2)}.\label{alpha}
\end{align}
Here, $2 a+b(p+1)>6$ is required for the convergence of the integral. Then, for the partially optically-thick part of the spectrum, the contribution of the part close to $z_0$ is negligible, and we can set $\xi_0$ above to zero. For the optically-thin part, $E/E_{\rm b}\gg 1$ [corresponding to $\tau_{\rm sa}\ll 1$, and thus $1-\exp(-\tau_{\rm sa})\simeq \tau_{\rm sa}$], we calculate the leading term in the series expansion in powers of $(E/E_{\rm b})^{-(p+4)/2}$ of equation (\ref{pl2}) and set $\xi_0=1$. This yields,
\begin{align}
&F_{\rm S}(E)\simeq \left(m_{\rm e} c\over h\right)^3 {(1+z_{\rm r})c \upi C_1(p) z_0^2\delta^3 B_{\rm cr}^{1/2}\epsilon_{\rm b}^{5/2} \tan\Theta\sin i\over 3 C_2(p) B_0^{1/2} D_L^2} \times\nonumber\\
&\qquad \begin{cases} \displaystyle{\Gamma_{\rm E}\left[\frac{2 a-6+b(p+1)}{2 a-2+b(p+2)}\right]}\frac{(\epsilon/\epsilon_{\rm b})^{\alpha}}{4+b}
 , &\epsilon\ll \epsilon_{\rm b};\cr
 \displaystyle{\frac{(\epsilon/\epsilon_{\rm b})^{(1-p)/2}}{2 a-6+b(p+1)}}, &\epsilon\gg \epsilon_{\rm b},\cr\end{cases}
\label{solution}
\end{align}
where $\Gamma_{\rm E}$ is the Euler Gamma function. We note that the value of $\alpha$ of equation (\ref{alpha}) agrees with that given by $\alpha_{\rm s1}$ of \citet{konigl81}. If we know $\alpha$ and $p$, we can invert this expression for $\alpha$. At $b=1$, we have $a=(10+\alpha p)/(5-2\alpha)$, and $a<2$ corresponds to $\alpha<0$. Note that the intersection of the two power laws is not exactly at $\epsilon_{\rm b}$. We also note that the argument of $\Gamma_{\rm E}(u)$ above is $0<u<1$, in which range $\Gamma_{\rm E}(u)\simeq 1/u$ [because $\Gamma_{\rm E}\simeq 1$ between 1 and 2], which provides an even simpler approximation. The condition for $\xi_0=1$ is $B_0>\epsilon B_{\rm cr}/\gamma_0^2$. Then, the low-energy partially-self-absorbed spectrum will become harder than $E^\alpha$ power law at $\epsilon<(B_0/B_{\rm cr})\gamma_0^2$. 

\label{lastpage}

\end{document}